**TITLE AND AUTHOR INFO**

Community-driven and water quality indicators of sanitation system failures in a rural U.S. community

Lorelay Mendoza Grijalva[1,#], Allisa G. Hastie[1,#], Meili Gong[1], Brenda Rojas Cala[1], Brandon Hunter[2,3], Stephanie Wallace[2], Rojelio Mejia[4], Catherine Flowers[2], Khalid K Osman[1,*], William A. Tarpeh[1,5,*]

1. Department of Civil and Environmental Engineering, Stanford University, Stanford, CA 94305, USA
2. Center for Rural Enterprise & Environmental Justice, Madison, AL, USA
3. Center for Diverse Leadership in Science, UCLA, Los Angeles, CA, USA
4. Baylor College of Medicine, Houston, TX 77030, USA
5. Department of Chemical Engineering, Stanford University, Stanford, CA 94305, USA

\# denotes equal contribution

\* Corresponding authors,

Email: wtarpeh@stanford.edu. Address: 443 Via Ortega, Room 387, Stanford, CA, 94305.

Telephone: 650-497-1324

Email: osmank@stanford.edu. Address: 473 Via Ortega, Stanford, CA, 94305.

**ABSTRACT**

Safe sanitation access is commonly believed to be ubiquitous in high-income countries; however, researchers and community advocates have exposed a glaring lack of access for many low- income communities and communities of color across the U.S. While this disparity has been identified and quantified at a high level, local and household-level implications of sanitation failures remain ill-defined. We develop a set of user-based and environmental measures to assess the performance of centralized wastewater systems, septic systems, and straight-piped systems in Lowndes County, Alabama. We combine qualitative, survey, and environmental sampled data to holistically compare user experiences across infrastructure types. This integrated approach reveals new routes of exposure to wastewater through informal household maintenance and system backups and provides evidence for the spread of wastewater-like contamination throughout the community. This work elucidates the severity of sanitation failures in one rural U.S. community and provides a framework to assess sanitation quality in other contexts with limited sanitation access in high-income countries.

**MAIN TEXT**

**1. INTRODUCTION**

The United Nations Sustainable Development Goal (SDG) 6 aims to achieve clean water and sanitation for all by 2030. The Joint Monitoring Programme for Water Supply, Sanitation and Hygiene estimates that 97% of the U.S. population has access to improved sanitation facilities, defined as locations that hygienically separate human waste from contact, enable safe disposal or transport from households, and are unique to each household (i.e., not shared).[1,2] As 2030 approaches, what stories remain untold by the 3% who do not have access? Overlooking those households can neglect public health risks for those individuals and their communities at large. Sanitation system shortcomings can take many forms, including discharge violations in centralized wastewater treatment facilities, improperly functioning household septic systems, and direct discharge of untreated household wastewater known as "straight piping". The burden of inadequate sanitation across the U.S. rests predominantly on low-income people and people of color, and does not appear to be purely the result of rurality, suggesting systemic drivers.[3,4,5]

Our specific study is rooted in Lowndes County in central Alabama in the region known as the "Black Belt". The Black Belt region is of particular interest for several reasons. First, its dark clay soil is largely impermeable, giving it limited suitability for septic systems.[6] This region is also an important case study because of its legacy of racial oppression and economic depression. These environmental and social conditions exhibit far-reaching effects in these communities, including that many households lack safely managed sanitation. Poor literacy, reconstruction era property laws, limited opportunities to build generational wealth, and criminalization of poverty have contributed to poor sanitation in Lowndes County.[7] Local residents and community organizations like the Center for Rural Enterprise and Environmental Justice (CREEJ) have advocated to correct these injustices and brought national policy and research attention to sanitation inequities.[8–10] Their advocacy work has increased the visibility of these issues on a national scale and drawn researchers from many backgrounds to the region.

Studies on sanitation inequities have revealed the prevalence of sanitation failures; however, extant key research gaps remain. Previous works have examined the relationship between sewage exposure and health effects to describe the public health implications of sanitation failures in rural U.S. communities. However, they explored a limited set of exposure pathways at the household level.[11] [12] [13] [14] From an infrastructural perspective, researchers have quantified sufficient sanitation based on traditional engineering measures, relying heavily on publicly available data sources.[6] [15] [16] Researchers have also examined these failures from a social science perspective, utilizing qualitative stakeholder data to identify barriers (e.g., lack of certified wastewater professionals, historic property laws) to improved sanitation services and infrastructure.[14] [17] Despite the growing body of literature on safe sanitation disparities in high-income countries, opportunities remain to integrate social science and engineering measures that go beyond enumerating infrastructure and documenting water quality. Traditional public health and engineering approaches do not adequately capture the lived experience at the household level, nor the larger community and environmental harm caused by failing systems.

One lingering point of debate is the effectiveness of extending centralized sewage system connection to residents. Some have proposed centralization (coupled with improved management and resource allocation) as an effective means of addressing these public health concerns,[15] while others have questioned the assumption that approaches typical to suburban areas will be effective in low-income rural contexts.[14] Traditional metrics like the sanitation ladder (five-level SDG ranking of systems: open defecation, unimproved, limited, basic, safely managed) would favor septic systems over straight piping, and centralized systems over septic as suitable solutions for these residents (SI Figure S1).[18] We do not directly contend with this classification, but propose a more nuanced and multifaceted exploration of different levels of service from household sanitation systems. Classifying systems beyond the SDG sanitation ladder can reveal inequities in high-income countries that have widespread access to safely managed systems, but persistent differences in the quality of sanitation services that households receive.

Overall, this study broadens the lens through which sanitation infrastructure performance is evaluated. We combine environmental water quality data with the richness of lived experiences via household-level surveys and interviews to assess different dimensions of sanitation function. This study uses a mixed-methods, community-based participatory research (CBPR) approach, merging both qualitative and quantitative analyses. From project conception to publication, community members were integrated into this study, beginning with open-ended engagement to allow residents' concerns and priorities to shape the research design. The mixed-methods approach allowed us to combine traditional quantitative engineering metrics with qualitative and quantitative resident-based measures of system performance. This study describes (1) how sanitation failures are experienced by residents across system types and (2) the potential impacts of these failures at household, community, and ecosystem scales. Our study is the first to measure and report different modes of system failures defined by both user-based metrics and environmental quality measures. Ultimately, elucidating the relationships between people and their infrastructure can enable informed interventions to address sanitation access in high-income countries in a holistic and inclusive manner.

## 2. MAIN TEXT

### 2.1 Residents face diverse types of sanitation shortcomings

We first analyzed households based on whether they had an installed wastewater treatment system. Among the 96 survey respondents, 91% (87 responses) reported having a formal system (i.e., septic or centralized), while 9% (9 responses) indicated the use of an informal system such as straight piping. Of those with a formal sanitation system, 41 are connected to a centralized wastewater system (towns of Fort Deposit, White Hall, or Hayneville) and 46 use a septic system (often in more remote parts of Lowndes County). We find that race and household size are largely consistent across system types with straight piping users reporting lower household income and households connected to city systems renting their homes at greater frequency (SI Table S1). Our study population is not representative of the average household in Lowndes County but rather captures the experience of lower income racial minorities in the community. Some of the challenges of validating these results are described in the limitations section.

From focus group discussions with residents, we developed operational definitions for several system performance indicators that capture resident experiences with and potential health risks of their sanitation systems (Table 1). System performance with respect to these indicators was captured through survey data and field sampling and observation. Together these metrics provide a nuanced understanding of sanitation system performance and potential household level exposure to septage. When examining these systems, we consider the infrastructure itself, its interactions with the natural environment, and the user perspective on the infrastructure.

Table 1. Sanitation infrastructure failure and performance metrics

| Metric | Definition | Data source | Purpose | System Performance | Exposure |
|---|---|---|---|---|---|
| Full backups frequency (annual) | Bubbling observed and heard in the system | Resident surveys | Indicator of system failure and severity, because pooling wastewater inside the home can lead to direct exposure to contaminants. | X | X |
| Partial backups frequency (annual) | Bubbling either observed or heard, but not both | Resident surveys | Indicator of partial system failure and severity, because slow drainage or bubbling may not result in direct exposure to contaminants. | X | |
| Self maintenance frequency (annual) | Frequency with which residents perform maintenance themselves | Resident surveys | Assesses resident involvement in upkeep (and perhaps prevents backups that | | X |

| | | | would otherwise occur) | | |
|---|---|---|---|---|---|
| Professional maintenance frequency (annual) | Annual frequency of professional servicing (e.g., pumping) | Resident surveys | Indicator of maintenance practices for septic systems and potential correlation with response to failure (backups) | X | |
| System satisfaction | Residents' overall satisfaction with their sanitation system | Resident surveys | Gauges perceived adequacy or effectiveness of system | X | |
| Standing water quality | If standing water was present on the property during the site visit, it was sampled and analyzed for wastewater | Field sampling and laboratory analysis | Indicator of failure severity, or exposure severity by seeing if water has contaminants and how similar it is to wastewater | X | X |

|   | characteristics (e.g., fecal viruses, COD) |   |   |   |   |
|---|---|---|---|---|---|

### 2.1.1 Performance can be assessed using resident perspectives and experiences

We measured performance of each sanitation system based on how consistently it removes waste from the home, its required maintenance, and overall satisfaction. We did not find statistically significant differences in performance between the three system types in any of these categories, but small effect sizes do exist (Table 2). Straight piping systems performed worse than city or septic systems with respect to backup frequency and satisfaction, but due to a small and unbalanced sample size, we cannot claim with sufficient confidence that there is a statistically significant difference between these systems. These measures of household performance are described in more detail in the following sections in addition to potential routes of exposure at the household level. We then examine how these indicators interact to provide novel insights into household sanitations systems within Lowndes County.

Table 2. Median value for performance metrics and result of Kruskall-Wallis test comparing system types. Mean values are reported in SI Table S2.

|   | Survey Respondents | | | | Effect size ($\eta^2$) | p-value* |
|---|---|---|---|---|---|---|
|   | Overall | City Sewage | Septic Users | Straight Piping | | |
| Respondents | 96 | 41 | 46 | 9 | | |
| Full Backup | Once | Once | 0-1 times | 2-4 times | 0.002 | 0.33 |
| Frequency (n = 96) | | | | | | |
| Partial Backup Frequency (n = 93) | Once | Once | Once | Once | -0.008 | 0.54 |
| Professional Maintenance (septic systems only: n = 44) | NA | NA | Never | NA | NA | NA |
| Informal Maintenance Frequency (n = 90) | Never | Never | Never | Never | 0.064 | 0.02 |
| System Satisfaction (n = 93) | Somewhat Satisfied | Somewhat Satisfied | Somewhat Satisfied | Somewhat Dissatisfied/ Neutral | 0.026 | 0.11 |

*p-values calculated from Kruskal-Wallis test comparing across all three system types. Critical p-value is 0.01 for a significance level of 5% after using the Sidak Correction[19] for multiple hypothesis testing

NA values for system types where metric does not apply

Effect size is reported as η². Small effect size: η² < 0.06, moderate effect size: 0.06 < η² < 0.14

*Backup frequency reveals inadequacies across system types*

The presence of wastewater inside homes indicates that sanitation systems are not functioning as intended. We advance on this indicator by measuring the reported *frequency and severity* of failures on an annual scale. We found that 67% (64 responses) of households experience some type of backups in their home (28 city systems, 28 septic systems, and 8 straight piping) (Figure 1). The largest portion of homes (45%, 43 responses) experienced both full and partial backups (18 city systems, 19 septic systems, 6 straight piping). Of those 43 homes, 11 reported full backups more than once a month over the last year. The frequency of these backups helps reveal persistent exposure to wastewater and wastewater constituents (e.g., pathogens, chemical contaminants) within homes.

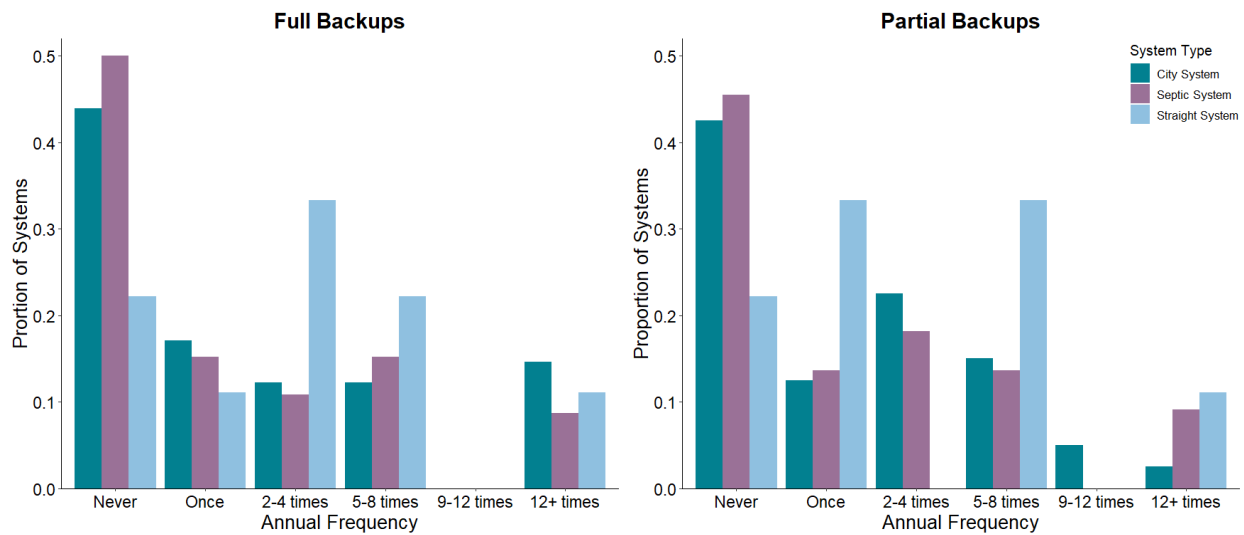

**Figure 1**. Annual count of full and partial system backups sorted by sanitation system type. Across all three system types, residents reported partial and full backups with a higher prevalence of partial system backups.

While installation of new septic systems or centralization are often proposed solutions to the sanitation crisis in the Black Belt, these systems also exhibit regular backups, causing stress and inconvenience for residents, as documented by an interview with a septic system user.

> *"But in that bathroom, yes, it has backed up in there, and I could hear it bubbling up in the stool, and the water in the sink wouldn't go down. And let me tell you something. You don't want that. That's a terrible feeling." [Interviewee 3] (SI Table S3)*

Examining multiple measures of system performance can enable detailed analysis of the quality of sanitation systems, which includes users' day-to-day experience with infrastructure and moves beyond categorizing the presence/absence of infrastructure..

*Maintenance practices have been taken on by residents themselves*

Regular maintenance is an essential practice for the long-term functioning of septic systems. Typical maintenance includes pumping sludge from the septic tank, inspecting the system, and replacing any damaged components. Generally, residential septic tanks are recommended to be pumped (i.e., emptied) every 3-5 years.[20][21] Many residents in Lowndes reported needing to have their systems serviced much more frequently. Of those households with septic systems, 24% reported having their tank pumped by a professional service at least twice in the last year, up to ten times more frequently than generally recommended (SI Figure S2). From our survey, residents ranked price (38% of respondents) and reputation (27% of respondents) as the largest factors that influence their choice of company servicing their system.

Residents also reported relying on informal maintenance practices. A quarter of surveyed residents, including 24% of those connected to city systems, reported performing their own maintenance (SI Figure S2). Anecdotally, households described a variety of maintenance practices including pumping their own tank and discharging it on their property (Focus Group Participant 1, 2, 7), repairing the subsurface field lines (Focus Group Participant 7) that distribute wastewater to the soil for filtration, and other *ad hoc* alterations (Interviewee 3). The effectiveness of these informal maintenance practices in preventing system backups is unclear because there was no statistically significant correlation between informal maintenance and backup frequency (Pearson's rho = 0.28, p = 0.06). These practices can have negative environmental effects by discharging pumped septage directly into the environment. In focus groups, personal maintenance arose as an example of self-sufficiency and problem-solving for urgent issues. Understanding this practice can help reveal tradeoffs between immediate household sanitation needs and long-term environmental protection.

> *"I think everybody that can probably just take and just go buy a septic pump and just knowing that we can just pump it ourselves and just do what we have to do…* [Focus group participant #7]
>
> *Anything to get it out, keep it from coming in the house."* [Focus group participant #9]

*Satisfaction is primarily driven by backup frequency*

Another measure of how well a sanitation system is performing its function is user satisfaction. On the median, straight piping households reported lower satisfaction, but city and septic systems were perceived equally ($\eta^2$=0.026, p=0.11) (Table S2). Resident interviews and survey responses revealed a high degree of nuance in their satisfaction with their system. For septic systems specifically, satisfaction was significantly correlated with the frequency of full backups (p < 0.001, rho = -0.43), but not service frequency (p=0.3). Alternatively, residents connected to centralized systems may be dissatisfied based on administrative and financial barriers. One participant shared that their city wastewater bill is set equal to their monthly drinking water bill. However, they reported an unusually high water bill (believed to be due to a leak), totaling $600 per month ($300 for water and $300 for wastewater). This example reveals how hidden issues with interconnected infrastructure systems can contribute to low satisfaction. This situation also highlights the financial burden and uncertainties around maintenance that residents face, even when using centralized systems.

**2.1.2 Performance can also be assessed through water quality**

*Standing water on private residences resembles wastewater*

One of the most apparent modes of sanitation failures in the Black Belt region is the pooling of wastewater in residents backyards from leaking septic tanks or straight pipe discharge. Among the 96 households we surveyed, we conducted field site visits at 11 households. Of those homes, 6 properties had visible standing water from which we collected samples to determine potential septage contamination. We classified standing water in terms of its similarity to municipal wastewater by analyzing the presence of fecal viruses (full summary statistics can be found in SI Table S4) and categorizing chemical oxygen demand (COD) levels into low, medium, and high ranges (SI Table S5, S6).

Five out of the six homes visted (83%) tested positive for microbial source tracker tomato brown rugose fruit virus (ToBRFV); of those five sites, four (67%) tested positive for pepper mild mottle virus (PMMoV, well-established indicator), and the same four sites also tested positive for human norovirus. The mean concentrations across all samples from private residences were 8.60 x $10^6$ gc/L for ToBRFV (median 2.17 x $10^5$ gc/L ), 3.68 x $10^5$ gc/L for PMMoV (median 2.26 x $10^5$ gc/L) , and 1.18 x $10^4$ gc/L for human norovirus (median 7.84 x $10^3$ gc/L); all three concentrations are similar to municipal wastewater levels (Figure 2A).[22] In addition to viruses, the water samples exhibited elevated chemical oxygen demand (COD) values, ranging from 7 mg/L to 1253 mg/L, where >550 mg/L is indicative of high-strength wastewater (Figure 2B). The mean COD value across all samples from private residences was 272.6 mg/L (median 168 mg/L), categorized as low-strength wastewater (SI Table S6). Sampling took place during a period of intermittent rainfall that diluted several samples, COD levels may have been even higher during a drier period.

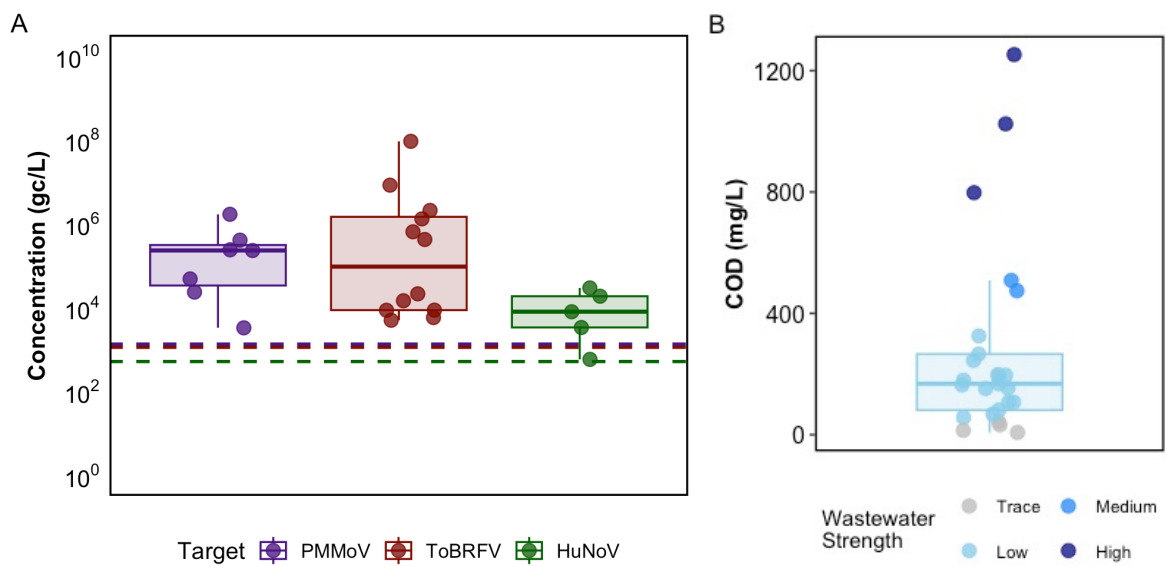

**Figure 2.** Virus RNA concentrations (gene copies per liter) (A) and chemical oxygen demand levels (mg/L) (B) in water samples from private residences. For Figure 2A, the dashed

horizontal lines represent the detection limits for each assay. PMMoV= pepper mild mottle virus (indicator), ToBRFV = tomato brown rugose fruit virus (indicator), HuNOV= human norovirus (pathogen). For Figure 2B, the COD wastewater strength was derived from values reported in wastewater treatment engineering literature[23–25] classified into trace (gray; <50 mg/L), low (50-350 mg/L) , medium (350-550) , and high strength wastewater (> 550 mg/L). All boxplots illustrate the distribution of each respective measurement. The central line within each boxplot represents the median (50th percentile), while the lower and upper edges correspond to the first (Q1, 25th percentile) and third quartiles (Q3, 75th percentile), respectively. The length of the box indicates the interquartile range (IQR = Q3 - Q1), and the whiskers extend to the smallest and largest values within 1.5 x IQR.

Overall, the water quality data strongly suggest septage contamination in sampled standing water. The contaminant concentration levels in the samples approach levels similar to untreated wastewater. The presence of viral pathogens coupled with the frequency of standing water on resident properties poses a potential health threat to residents. About 30%of survey respondents (29 respondents) reported standing water on their property (attributed to rainfall or creek overflows) at least once a month (SI Fig S3). While we cannot confirm that all reported instances of standing water contain septage, Lowndes receives rainfall year-round and the regularity of standing water events can mobilize contaminants and increase exposure for both residents with failing sanitation infrastructure and their neighbors.

*Describing household contaminant exposure*

Through our qualitative data collection, surveys, and environmental sampling we confirmed the presence of septage on private properties and described exposure through backups and informal maintenance. The need for informal maintenance is not only an indicator of system failure, but also presents an additional exposure pathway that has not previously been documented among septic users. Our study is the first to explore and quantify the prevalence and frequency of self-maintenance behaviors and identify these practices as an exposure pathway. This work also confirms the results of previous studies that have identified standing septage on private properties in Lowndes County and the presence of raw sewage in the home.

> *"Every time you out there it's exposure, you never know what's out there. What you're breathing in, what your skin's coming in contact with." [Focus group participant #2]*

Residents have indicated an awareness of the risks they face through multiple routes of exposure, but have lingering questions on the specific contaminants with which they come into contact. This study generates specific results on the exposures that residents face at the household scale, generating actionable insights for participants and the community at-large.

### 2.1.3 Relationships between user experience and sanitation performance are complex

Across operational measures of system performance, we find a variety of experiences within and between system types. While some general trends appear, we also find several insightful exceptions and nuances. Figure 3A indicates that across system types, high rates of satisfaction are inversely correlated with observed backup frequency ($p = 0.00001$, $rho = -0.43$) indicating that consistent removal of waste from the home drives satisfaction. In contrast, maintenance frequency is not significantly correlated to satisfaction, which is consistent with residents' willingness to perform maintenance and make upgrades to their systems. Combining user-reported data with water quality data helps cross-validate potential indicators and reveals intersectional insights that neither dataset alone can provide (Fig 3B). Even with regular formal and informal maintenance, residents experienced standing water consistent with septage on their property (Figure 3B). Based on these performance measures, we highlight two archetypes of households that offer nuances to the general trends.

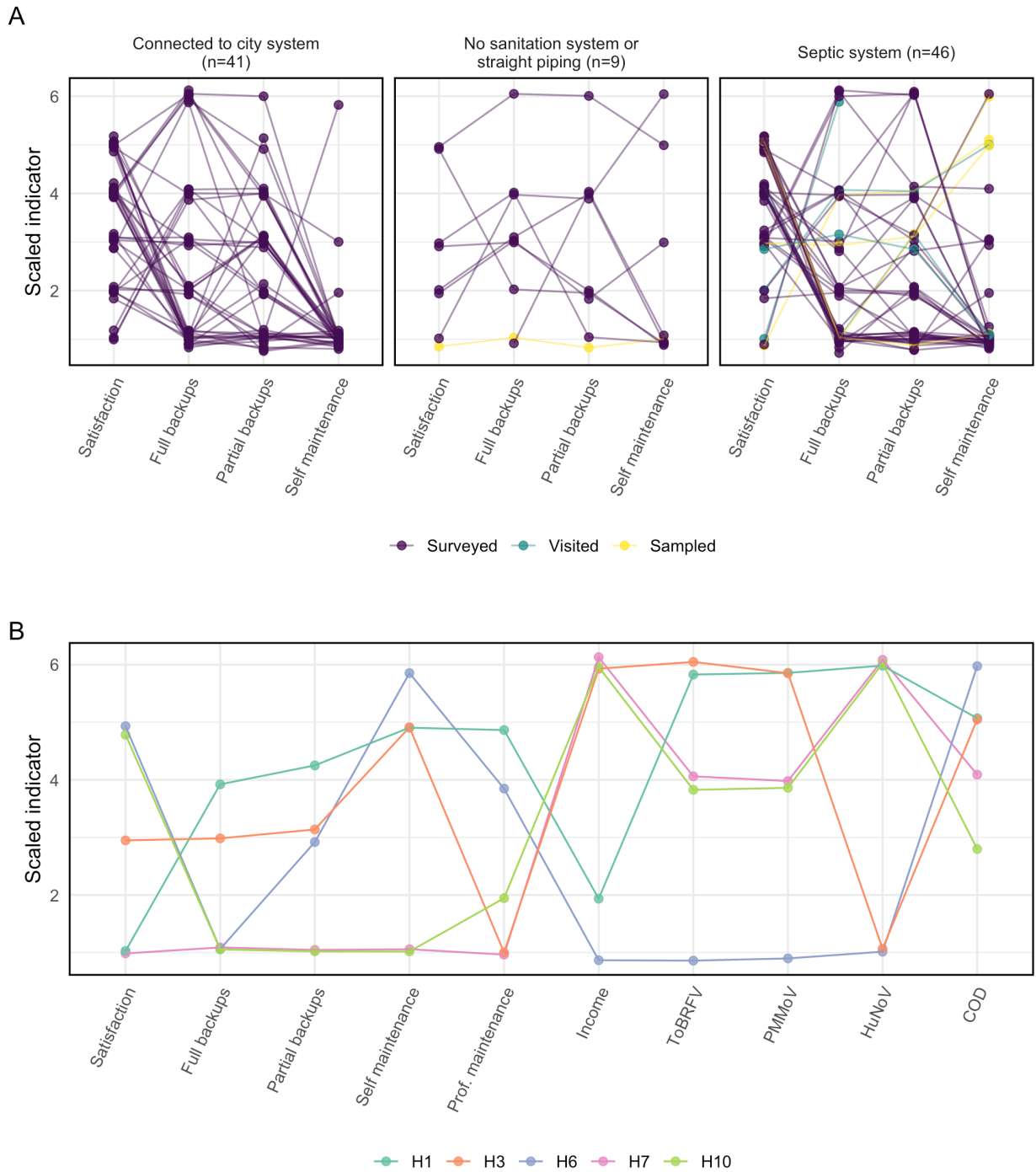

**Figure 3.** Parallel axis plots showing survey responses about system performance and water quality data. The most common experiences (measured with scaled indicators, Table S7) exhibit

the largest number of overlapping points and lines, while allowing deconvolution of user-reported and water quality data for each household. Figure 3A shows satisfaction and performance for all survey respondents. Violet lines are a baseline color for all survey respondents. The other colored lines denote homes where we also visited and sampled standing water; "visited" homes are a subset of "surveyed" homes, and "sampled" homes are a subset of "visited" homes. Figure 3B reflects the homes visited where samples were taken and water quality analyzed. Points on the y-axis are on a whole number scale, where higher point values on the scale represent higher values in survey responses and water quality results. SI Table S7 details the scale used on the y-axis in all plots. Data points are slightly jittered horizontally to reduce overlap and improve visualization.

*Residents prioritize a system that works*

Household 6 (blue line on Figure 3B) reported a high level of satisfaction with their system despite frequent personal and professional maintenance. Standing water exhibited high levels of COD but virus loads below detection, indicating that the system serves its environmental function relatively well, but residents still experience some partial backups in the home. Although household 6 reported that their system required nine maintenance sessions per year (a combination of personal and professional), they reported high satisfaction with their system. Some individuals value their system functionality so much that even a high burden of maintenance does not lower their satisfaction. What matters most is that the system fulfills its intended purpose. Satisfaction aside, this household exemplifies the high degree of effort required for septic systems to function even moderately well in the conditions of Lowndes County.

*Low maintenance burdens are necessary, but not sufficient*

Households 7 and 10 both display low rates of backups in the home, one indicator of acceptable performance (Figure 3B). Both households also report minimal professional and informal maintenance. Despite similar water quality levels, these homes differ the most in satisfaction. Household 10, which has a septic system, reported being very satisfied with their system, likely due to the perception that the system is functioning as intended and requires no action. Household 7, which relies on straight piping, reported extremely low satisfaction, despite also having no maintenance burdens. This contrast suggests that satisfaction is shaped not only by the day-to-day burdens, but also by residents' expectations for adequate and dignified infrastructure. Even in the absence of pipe backups and maintenance responsibilities, residents may still feel dissatisfied with informal solutions that fall short of what they consider proper sanitation.

## 2.2 Impacts of sanitation failures in the community and environment

### 2.2.1 Failing sanitation systems impact the surrounding environment

Failing sanitation systems pose environmental threats by dispersing wastewater contaminants beyond their point sources (e.g., individual homes with failing systems) into the broader environment. To quantify potential exposure beyond households, we sampled from seven environmental sites including parks, reservoirs, and recreational areas to assess the distribution of wastewater beyond the home. Three additional sampling sites were public locations – two schools (Sites S1-S2) and a public park (Site S3) – that may have a sanitation system and were not private residences.

Environmental samples tested positive for human fecal viruses, indicating potential contamination from wastewater. ToBRFV was detected at all sites, PMMoV was detected at four of seven environmental sites, and norovirus was detected at one of seven environmental sites (Figure 4A). Abundance followed the same trend as prevalence (ToBRFV highest, followed by PMMoV and norovirus). Site E1, where norovirus was detected, was a culvert that channeled water underneath a road (Figure 4C). Although all of the recreational sites were along the Alabama river (E2, E5, and E6) and had detectable levels of fecal markers ToBRFV and/or PMMoV, none exhibited detectable levels of the human pathogen norovirus.

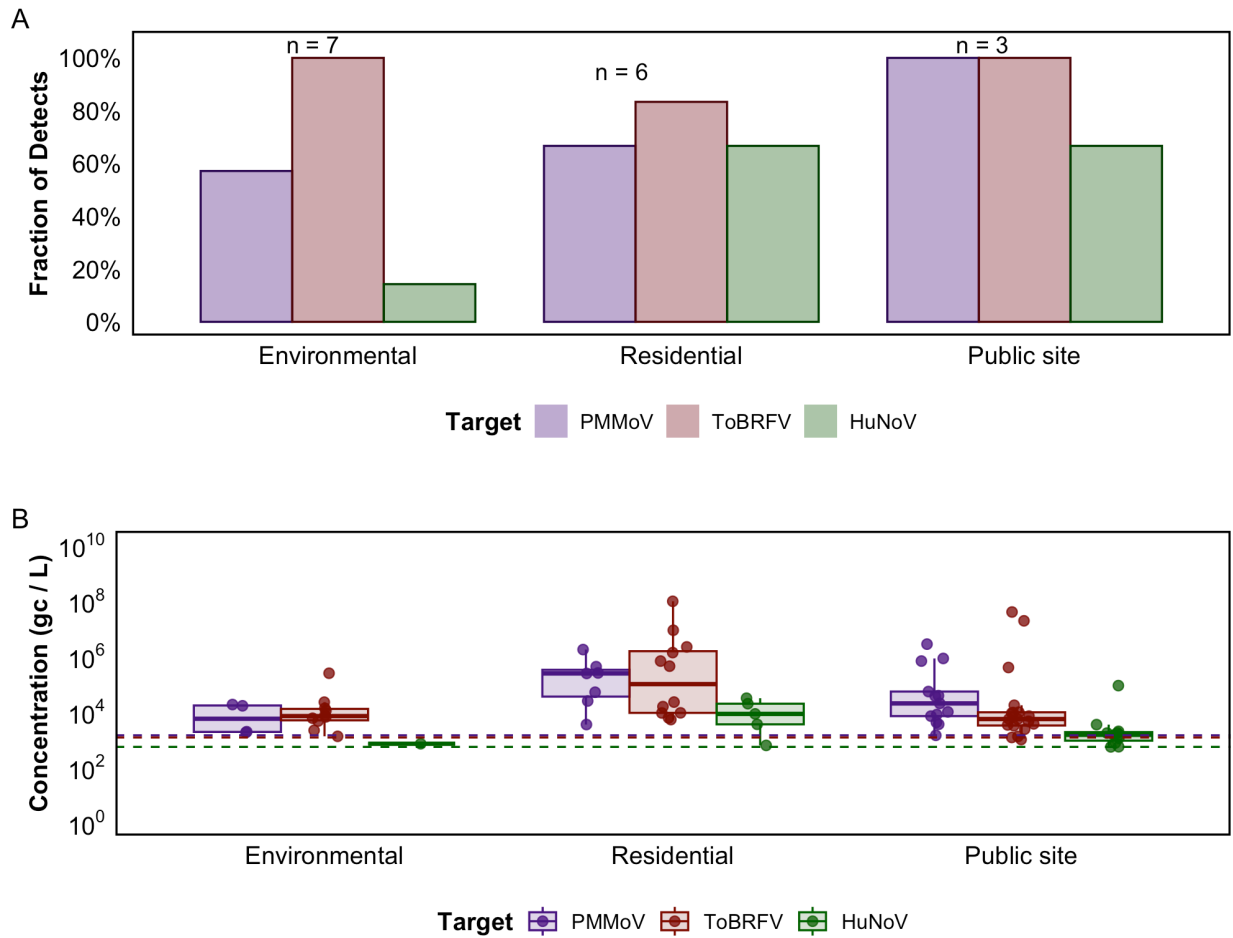

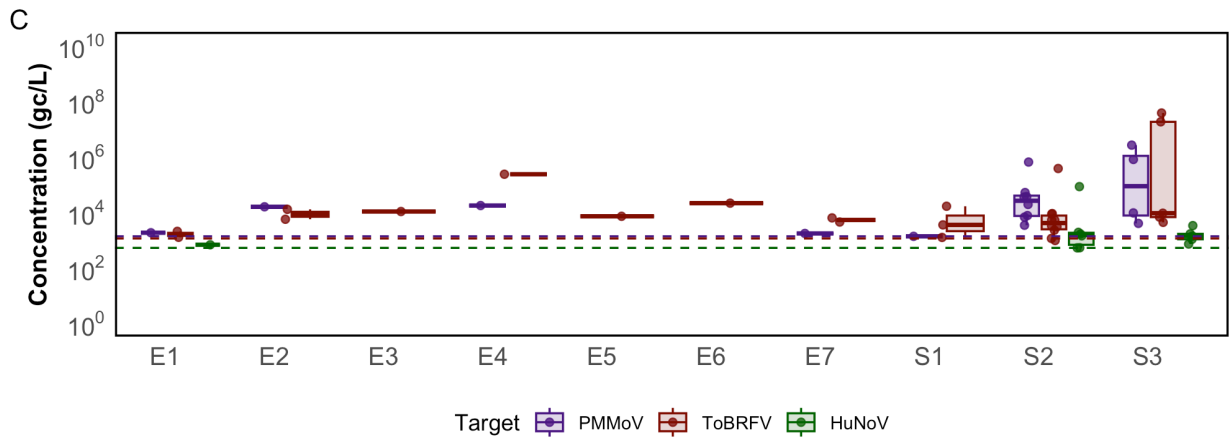

**Figure 4** (A) Bar plot illustrating the fraction of unique sites within each category where specific viruses were detected. For instance, among the 7 unique environmental sites, 57% (4 out of 7) tested positive for PMMoV, 100% for ToBRFV, and only one site was positive for norovirus. (B) Box plots showing the distribution of virus concentrations across different site categories. (C) Virus RNA concentrations for each site that was not a home, including all environmental sites (E), the two schools (S1 and S2), and the public park (S3). The log-scaled y-axis highlights the range and variability in concentrations detected for each virus. All boxplots illustrate the distribution of each respective measurement. The central line within each boxplot represents the median (50th percentile), while the solid data point inside represents the mean. The lower and upper edges of the box correspond to the first (Q1, 25th percentile) and third quartiles (Q3, 75th percentile), respectively. The length of the box indicates the interquartile range (IQR = Q3 - Q1), and the whiskers extend to the smallest and largest values within 1.5 x IQR.

In general, environmental samples (E1-E7 in Figure 4C) had lower concentrations than public sites (S1-S3). The mean concentration across all environmental sites was $3.00 \times 10^4$ gc/L (median $6.55 \times 10^3$ gc/L) for ToBRFV, $8.98 \times 10^3$ gc/L for PMMoV (median $8.58 \times 10^3$ gc/L), and $6.56 \times 10^2$ gc/L for norovirus (median $6.56 \times 10^2$ gc/L) (Figure 5B). Public sites S1 and S2 were public schools, and S3 a public park. Both S2 and S3 had wastewater infrastructure onsite, which may explain the higher virus concentration and the detection of all three viruses (Figure 4A and Figure 4C). These results suggest they can serve as hotspots for wastewater-associated contamination. ToBRFV and PMMoV were detected at all three of the public sites we sampled, and norovirus was detected at two of them. The mean concentrations in public site samples were $3.09 \times 10^6$ gc/L for ToBRFV (median $5.11 \times 10^3$ gc/L), $3.21 \times 10^5$ gc/L for PMMoV (median $1.87 \times 10^4$ gc/L), $8.73 \times 10^3$ gc/L for norovirus (median $1.38 \times 10^3$ gc/L) (Figure 4B). Only ToBRFV concentrations were significantly different across site types ($H = 8.35$ $p = 0.01$).

We also assessed environmental sites for wastewater contamination via chemical oxygen demand and nutrient concentrations. Nitrogen and phosphorus concentrations are regulated contaminants and are required to be removed by wastewater treatment plants.[26] The mean COD value for environmental samples was 33.7 mg/L (Figure 5A), which is slightly higher than normal surface waters (like rivers and ponds). Private residences had significantly higher COD levels (mean 272.6 mg/L, median 168 mg/L) compared to both environmental sites ($p < 0.001$) and public sites ($p = 0.002$). There was no significant difference between environmental and public sites. Private residences had the highest nutrient levels overall, especially for ammonium and phosphate (mean 4.33 mg/L and 6.32 mg/L, respectively) (Figure 5B). Public sites had moderate nutrient levels, and environmental sites had the lowest nutrient concentrations, with no detectable levels for nitrate, nitrite, and phosphate. Across all nutrients, median concentrations were consistently 0, but occasionally spiked in certain samples, and differences across sites were not significantly different (Figure 5B).

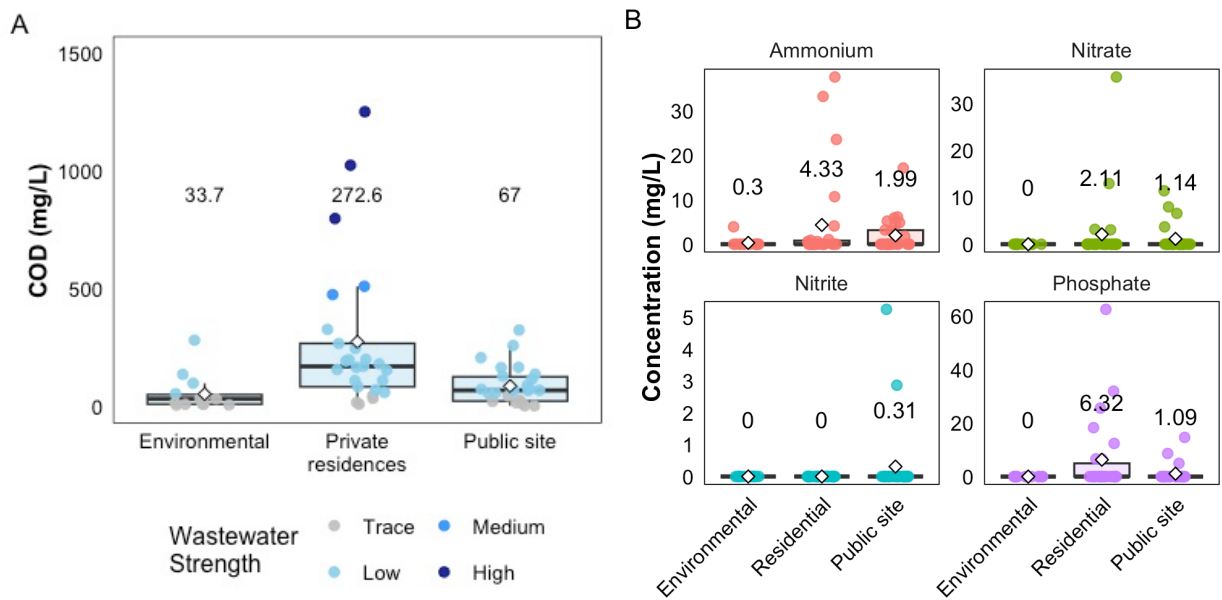

**Figure 5.** (A) Chemical oxygen demand and (B) nutrient concentrations in samples at different locations. Annotated values shown are mean values for each type of site. Wastewater strength values were constructed from values found in the literature (SI Table S6).

Standing water from homes most resembled high-strength wastewater, followed by public sites, then environmental water samples. Despite the lower COD and nutrient values in some environmental samples (Figure 5B), fecal viruses were still detected at public and environmental sites, albeit at reduced concentrations compared to households (Figure 4B). Ultimately, failing sanitation systems at private residences exhibits the highest COD and nutrient values measured (Figure 5B), and thus can contribute to potential exposure to pathogens and chemical contaminants throughout the study community and environment.

Early estimates for soil suitability for onsite wastewater treatment systems classify much of the American Black Belt region as having unsuitable soil, mostly due to soil density and other properties.[6] For the same reason, larger-scale nature-based solutions like treatment lagoons can also be just as unsuitable, and sites in close proximity to these treatment systems can pose unexpected exposure risks.

To illustrate the potential community risks, we consider a local high school adjacent to a series of wastewater treatment ponds. The outfall of the last pond discharged into a tree line next to the football field. During our visit, light rainfall created a puddle between the treeline where the outfall was located and the football field. This sample had a COD concentration lower than the wastewater pond samples but higher than other environmental samples (Figure S4A). However, all three viruses, including norovirus, were detected (Figure S4B), further suggesting that while rainfall may dilute contaminants, it can also disperse them to areas where exposure is possible but often overlooked. Overall, our fieldwork results suggest that exposure to wastewater-related hazards across the community is likely underestimated. These insights would not have come to light without engaging with community members and conducting community-informed, mixed-methods site visits in Lowndes County.

**2.3 Limitations**

Although our study provides rich insights into one location, we recognize that a small sample size presents some limitations to the generalizability of the results. Our study participants are not demographically representative of the "average" household in the county but rather are lower income, more likely to be Black or African American, and more likely to rent their home (SI Table S1). Survey questions were developed based on best practices to ensure validity and reliability but there still exists some degree of uncertainty in self-reported data. We were also unable to confirm all the home addresses shared by residents because of the rural nature of the community with some streets having informal or historic names. Despite the small sample size, the depth of our sustained community engagement – through focus groups, co-developed surveys, in-depth interviews, and site visits – fostered a rich descriptive understanding of household conditions. Participants were recruited through a variety of methods: existing relationship with CREEJ, response to mailed invitations to community survey sessions, attendance at a local church, and through word-of-mouth from other participants. Given the sensitive nature of these issues, residents with severe sanitation challenges may have felt unwilling to share their experiences, or may have been biased towards participation to highlight their experiences. Alternatively, households where sanitation is a low-salience issue may have been less motivated to participate in this study.

Counter to common practice when reporting data from environmental samples, we do not provide a map of sampling locations to maintain participant privacy due to the sensitive nature of this work. Until recently, the Alabama Department of Public Health penalized households for non-compliant septic systems, which effectively criminalizes sanitation system failures.[10] A common limitation in environmental sampling using molecular methods is that droplet digital PCR (ddPCR) detects viral nucleic acids rather than infectious virus. This method confirms the presence and concentration of viral genetic material, but it does not distinguish between intact, infectious viruses and degraded, non-viable remnants. Thus the results cannot be assumed to directly measure transmission risk, but the presence of viral material is a necessary prerequisite for exposure.

## 2.4 Discussion

Historically, sanitation access has been assessed by counting infrastructure (e.g., septic tanks, toilets, treatment plants) rather than considering lived experiences. As a result, current estimates for sanitation access may overestimate how many people actually have access to safely managed sanitation.[27] Surveys that simply record the presence of septic tanks – assuming they function properly – may classify all septic tank owners as having safely managed sanitation. However, when resident perspectives are integrated into this assessment, a different and more nuanced picture emerges. Many homes experience frequent backups and standing water contaminated with wastewater, which indicates inadequate treatment and disposal. According to our paired survey responses on backup frequency and measured residential water quality, many of these households would be reclassified lower on the sanitation service ladder (e.g., limited or improved sanitation).

Using a -approach led us to develop survey questions about backups and maintenance in direct response to concerns raised in focus groups. As an outcome of centering residents' priorities, we introduce two quantitative user-reported indicators to evaluate sanitation access: (1) the frequency and severity of system failures and (2) the time and effort invested in maintenance. We defined failure severity based on resident descriptions (full and partial backups) and then measured their annual frequency. We also collected data on maintenance practices and frequency. Similar estimates have been made for drinking water by including time and quality-of-life indicators to measure safe drinking water access.[28–31] Researchers have recently begun to advocate for the use of analogous detailed indicators in the sanitation sector.[32,33] This study is the first to propose specific indicators in a high-income country, rooted in a research design that centers community voices. Our findings show that residents are having their systems serviced up to ten times more frequently than the EPA recommends, often taking on much of the work themselves. Capturing time and efforts expended by residents as key indicators can help determine whether new systems are performing as intended. As sanitation services improve, residents should experience reductions in maintenance frequency; furthermore, maintenance frequency could be a useful addition to the sanitation service ladder.

To our knowledge this is the first study in Lowndes to incorporate water sampling and analyze for viral and chemical markers of sewage contamination. Previous studies have sampled soil samples[11] and stool samples[12,13] and examined them for various enteric pathogens. Sampling soil matters because septage percolates through it, and some pathogens rely on it for part of their life cycle. Sampling stool is useful for confirming individual infections because pathogens are shed in feces. Sampling water is particularly important not only because of the well-established links between poor sanitation and waterborne disease, but also because precipitation will play a role in anticipating how extreme weather and climate change might affect sanitation-related risks. Heavy rainfall is common in Alabama, and even sewered areas elsewhere are prone to overflows.[34] The present study expands on previous work by incorporating water sampling, and including both microbial and chemical markers; we confirm that standing water on the properties is often contaminated and can be an exposure route to wastewater for residents.

Frequent personal maintenance of failing sanitation systems can become a risk factor for exposure to untreated septage. Consequently, exposure to waterborne contaminants among communities in underserved areas is likely underestimated. Once quantified, maintenance-associated exposures can be incorporated into microbial and chemical risk assessment modelling. These risk assessments have been conducted for residents and sanitation workers in informal settlements in low- and middle-income countries.[35] There may be additional insights to draw from residents in high-income countries who conduct self-maintenance, and thus dual roles as residents and informal sanitation workers. Residents try to reduce their exposure risk inside the home (from backups) by performing regular maintenance and pumping, but can experience a tradeoff between exposure due to maintenance and exposure due to backups and other system failures. Wastewater contaminants in environmental surface water and standing water at public locations indicate that sanitation failures also have community-wide public health impacts, and present additional exposure, even for residents with functional household systems. Sanitation systems can be point sources of wastewater contaminants that are mobilized away from the home by rainfall and creek flows. Ultimately, sanitation failures are a collective problem that require systemic solutions.

Sustainable solutions to solve sanitation problems require a nuanced approach and community input. Many residents, regardless of sanitation system type, face frequent failures and express dissatisfaction. They often need to manage tradeoffs between financial costs for professional maintenance, time and effort for personal maintenance, and hygiene concerns both in and out of the home. Overall, the absence of mixed-methods approaches in addressing sanitation disparities omits several key insights and priorities; extremely relevant user burdens can go unheard and unmeasured. While small in scale, this study introduces specific indicators for assessing sanitation system performances and reveals new exposure routes to untreated wastewater. Addressing the global sanitation crisis and achieving SDG 6 for all will require mixed-methods approaches for evaluating and benchmarking the effectiveness of implemented solutions.

3. **Methods**

We used qualitative data collected from community focus groups to develop a survey of infrastructure performance and sociodemographic data with additional in-depth individual interviews. To assess public and environmental health impacts, we conducted site visits to measure water quality across private residences, developed public sites, and the broader environment. Specifically, we measured microbial source tracking (MST) markers including an established indicator, pepper mild mottle virus (PMMoV); an emerging MST tomato brown rugose fruit virus (ToBRFV), and an enteric pathogenic virus (HuNoV). We also measured chemical oxygen demand (COD) and nutrient concentrations (nitrogen and phosphorus) to provide a holistic assessment of wastewater-derived contaminants. All human centered research methods were approved by Stanford University's Institutional Review Board (#67721) and WCG (#IRBOOOOO533).

**3.1 Community Based Participatory Research (CBPR) approach**

This study was a collaboration between academic researchers, leadership and staff at the Center for Rural Enterprise and Environmental Justice (CREEJ), and community residents and leaders in Lowndes County. Aligning with CBPR principles, our academic research group sought out well-connected community partners to ensure the accessibility and relevance of our work to community needs. Study leadership included expert community representation; Catherine Flowers, the founder of CREEJ, served as a co-principal investigator (co-PI) on this project. CREEJ is a non-profit organization dedicated to reversing health, economic, and environmental disparities related to insufficient wastewater infrastructure throughout Lowndes County. This collaboration involved shared decision-making authority and equitable access to project resources (e.g., grant funding). With access to grant funding, CREEJ was able to hire additional staff, pursue their own IRB approval through WCG, and support travel to academic conferences. Founded in 2002, CREEJ's deep relationships, engagement, and partnership with Lowndes County communities provided invaluable contextual on-the-ground information that shaped the research scope, design, and execution of this collaborative effort.

To ensure equitable data ownership and autonomy, CREEJ obtained their own institutional Review Board (IRB) approval, granting them full rights to all collected data, which posed institutional challenges due to their non-academic status. Negotiating this arrangement required significant advocacy but was necessary to ensure equity in resource distribution. A shared Google Drive was used to enable a collaborative platform with live access to all workflows. Open data access enabled all project leaders to review and contribute to working documents in real time. This approach fostered transparency and ensured that the community had continuous opportunities to engage with and influence the research process. Key members of our community partnership are acknowledged as merited authors for their expertise, thought leadership, and skilled labor.

We took several steps to maximize the accessibility to our study to community members. Prior to the development of the survey and field study, we conducted focus groups in Lowndes to build rapport and center community priorities. Before each session, we introduced ourselves, distributed printed and laminated materials for the participants to keep, and presented a slideshow to build rapport, invite feedback, and answer questions. In collaboration with our community partners, we determined the most convenient days and times for participants, aligning sessions with work schedules to maximize participation. We scheduled focus groups during weekday evenings at familiar, community-centered locations like the local town hall and church. We also provided refreshments and welcomed children, recognizing that some participants required childcare. We have planned a summary for Lowndes residents who participated or are interested in the findings. We will distribute the materials, return to the community to present the results, and address any questions. The timeline for the project including site visits is attached in SI Table S8.

**3.2 Mixed Methods**

*Study design*

In this study we used a mixed-methods approach defined as "research in which the investigator collects and analyzes data, integrates the findings, and draws inferences using both qualitative and quantitative data."[36] We leveraged an exploratory approach, with initial qualitative data being used to inform our subsequent collection of quantitative data. These data were integrated through an embedded approach where our primary research questions were answered using quantitative data and supported or further examined with qualitative analysis.[37] We ultimately combined data from four different sources: community focus groups, individual interviews, community surveys, and standing water samples (SI Table S9).

*Qualitative data collection*

Community members' perspectives and experiences with their sanitation systems were collected through focus groups and one-on-one semi-structured interviews. A total of 22 focus group participants were recruited by CREEJ and invited to participate in one of three 1-hour facilitated group discussions about sanitation inadequacies in their community (SI Table S10). Participants were compensated for their participation. Because of the community's engagement we expanded from our planned 2 focus groups to include a total of 3. Each focus group was conducted by one facilitator and one note taker. The primary questions asked in each focus group are provided in SI Table S11 with follow-up questions based on the discussion.

Qualitative coding of the focus groups took a hybrid inductive-deductive approach. After multiple rounds of revising the codebook and meeting between the two authors coding the transcripts, we achieved a kappa coefficient of 0.6, indicating sufficient inter-rater reliability for the purposes of this study.[38] Our final codebook with 5 major themes (adaptation, environment, health, infrastructure, and organization and systems) can be found in SI Table S12. During survey completion community members were also invited to participate in a one-on-one interview. The purpose of these interviews was to expand on the questions from the survey and gather more specific details on their individual household sanitation systems and goals and priorities for solutions. We conducted a total of six interviews with residents that were later coded with the same codebook developed from focus groups.

*Survey data*

From the key themes identified in the focus groups and our core research questions, the research team and community partners co-produced a survey for distribution across the community. Group input decided the language for each question and response options, order of questions, and length through four rounds of iterations. Community members were invited to complete the survey through two different pathways, either by coming to a "survey session" hosted in three central locations across the community or by the research team providing a survey while collecting water samples from their home. A total of 115 surveys were completed through a paper version or submitted through Qualtrics on a tablet provided by the research team. After processing and assessing for quality control (SI Figure S5), 96 surveys were included in the final dataset. The paper version of the survey is attached in SI S13.

*Field-based sample collection*

Over five consecutive days in February 2024, 59 unique water samples were collected across private homes, public land, and public waters in Lowndes County. See SI Tables S14 and S15 for subcategories of sample types and count. Private residences were selected for sampling based on the recommendations and relationships of community partners. These community partners sought to secure as many houses for visits as possible, utilizing their personal and professional networks. Eleven homes volunteered to have a site visit and water sampled on their premises. These homeowners may have had existing concerns about their sanitation systems. Nine of the eleven homes also filled out a survey, making them a subset of the survey respondents.

Our goal was to sample as many environmental sites as possible, focusing on locations where people gather and socialize for recreation. We first curated a list of potential sites using recreational locations identified on Google Maps. CREEJ partners reviewed the list and provided a curated, updated list for visits and sampling. We sampled 11 other environmental sites, one public park, and two public schools, resulting in a total of 14 non-residential sites. CREEJ partners served as site guides as the team stopped at each designated location exit and proceeded with sampling activities. Additional details on sample collections, transportation, and storage can be found in supplemental methods SI Methods S1.

**3.3 Analysis Methods**

*Survey Data Analysis*

We conducted statistical testing to determine differences in performance between city systems, septic systems, and straight piping based on backups frequency and satisfaction. We used the Kruskal-Wallis test to compare between all three system types and to determine if our sample data is taken from a population with the same median. Our null hypothesis is that the median value for each metric will be the same between system types. Using the Sidak correction and performing four comparisons (full backup frequency, partial backup frequency, informal maintenance frequency, resident satisfaction), our adjusted acceptable p-value was 0.01 for a significance level of 5% and statistical power of 0.8.[19]

Next, we examined whether backup frequency and frequency of informal pumping are correlated with overall satisfaction across system types using Spearman's rank order correlation and an adjusted alpha value of 0.025. We also ran tests to determine which factors, if any, are correlated with septic system performance, measured as satisfaction and backup frequency. We ran correlation tests on factors we expect to influence septic performance based on physical system constraints, household characteristics, and maintenance practices. These factors include soil suitability, frequency of standing water due to rainfall, frequency of standing water due to creek overflows, social vulnerability, system age, household size, frequency of professional system pumping, and frequency of informal system pumping. The adjusted alpha value for those tests was 0.006.

*Field Sample Analysis*

For the field samples collected, three categories of water quality indicators were measured and used to assess their similarity to wastewater: nutrients (nitrogen and phosphorus), chemical oxygen demand (COD), and viruses (summary in SI Table S16). Nutrients are commonly associated with anthropogenic sources, and their release into the environment can have negative impacts. These compounds are prevalent in wastewater, agricultural runoff, and other human-related sources. Specifically, four compounds were measured: total ammonia, nitrate, nitrite, and total phosphate. Chemical oxygen demand quantifies the amount of oxygen required to oxidize organic matter in a sample, and is an established indicator of wastewater quality. COD is widely used in wastewater analysis and serves as a chemical indicator of organic pollution. To contextualize the results, we developed a reference table based on literature values to classify wastewater strength and determine how similar the values were to wastewater (SI Table S6).

Three viruses were measured in the samples, including two microbial source tracker (MST) markers, and norovirus, an enteric pathogen. The microbial source trackers are viruses used to indicate fecal contamination from humans, but are not pathogenic themselves. The MST markers included PMMoV, a well-established indicator, and ToBRFV, an emerging marker that has also been found to be ubiquitous in human feces but has not yet been measured in Alabama. A description of the viral targets and justification can be found in SI Table S17.

Ion chromatography was also used to analyze water samples. In total, twelve ions were measured. While none of the measured ions alone are a definitive indicator of wastewater, IC analysis provided valuable information to better understand overall water composition (SI Figures S6-S8). Cations measured were ammonium ($NH_4^+$), calcium ($Ca_2^+$), lithium ($Li^+$), magnesium ($Mg_2^+$), potassium ($K^+$), sodium ($Na^+$). Anions were bromide, ($Br^-$), chloride ($Cl^-$), nitrate ($NO_3^-$), Nitrite ($NO_2^-$), Phosphate ($PO_4^{3-}$), and sulphate ($SO_4^{2-}$). Within 30 days of collection, samples were removed from refrigeration, filtered through a 0.22 um filter, and diluted as needed prior to ion chromatography analysis. Anion concentrations (i.e., fluoride, chloride, nitrite, bromide, nitrate, phosphate, and sulfate) were measured via anion chromatography (45 mM carbonate/1.4 mM bicarbonate eluent at 1.0 mL/min, 25 ul injection volume, and 26 mA suppressor system). Cation concentrations (lithium, sodium, ammonium, potassium, magnesium, calcium) were measured via cation chromatography (20 mM methanesulfonic acid eluent at 1.0 mL/min, IonPac CS12A column, 125 uL injection volume, 59 mA suppressor system) on a Thermo Fisher Scientific Dionex ICS-6000 system (Waltham, MA). Cation samples were acidified with 2 M sulfuric acid prior to analysis to ensure ammonia stability. Lower detection limits for each species can be found in SI Table S18. For both viruses and chemical contaminants, results were processed using R (version 2024.09.0+375) for data wrangling and data visualization. Statistical analyses were performed in GraphPad Prism version 10.3.1 (464). For viruses, non-detects were excluded in the analysis, thus reducing the sample size. For all water quality measurement data, nonparametric Kruskal-Wallis methods were used to compare medians between private residences, environmental sites, and public sites.

Sample pH and conductivity were measured onsite using field pH and conductivity probes. (SG68 Handheld pH meter, pH probe, conductivity probe, Mettler Toledo, Oakland, CA). Chemical oxygen demand (COD) was determined colorimetrically by Hach Method 8000 using high-range (0 – 1,500 mg/L) reagent kits (CHEMetrics, Midland, VA) with a DRB200 digital reactor block and a DR1900 portable spectrophotometer (Hach Company, Loveland, Colorado). We conducted triplicate COD measurements on a random subset (10%) of samples, and the average value was used for all reporting and downstream analyses for those samples (Table S19). All in-field analyses were initiated at the end of each collection day. Conductivity and pH were measured the same day as collection. COD measurements were also started on the same day whenever possible, but due to long processing times, some samples were completed the following day. Sample preservation in the field involved refrigeration at approximately 4 $^{\circ}$C. All COD analyses were completed within 72 hours of sample collection.

Digital droplet PCR (ddPCR) was used for virus detection. In brief, water samples were concentrated via centrifugation and RNA was extracted using Qiagen Allprep Powerviral Kits (Qiagen, Germantown, MD) following the manufacturer's instructions. The extracted RNA was then quantified using ddPCR (Biorad). Additional details on the assays and methodologies are provided in supplemental Tables S20, and supplemental methods M2.

To investigate whether wastewater contaminants originated from the septic tank or were ubiquitous to the environment, we conducted a small field experiment. When available, we collected water samples from standing water in the front of properties as a negative control and compared to back of property concentrations. The back-of-house samples tested positive for all three viruses while only two of the four front-of-house samples tested positive for ToBRFV (SI Fig S9A). Median ToBRFV RNA concentrations in back-of-house samples were two orders of magnitude higher than those in the front (SI Fig S9B). The negative control sample from the front of the house indicated that these viruses are not necessarily ubiquitous, demonstrating that detection is not always expected (SI Fig S9). Together, these results increase the confidence in the conclusion that contaminants present in standing water originate from untreated wastewater.

*Acknowledgments*

The authors acknowledge the Lowndes community members who participated in our focus groups, took our surveys, and opened up their homes to us for site visits and sample collection; without them, this study would not have been possible. We also acknowledge Dylan Creamer for their assistance with data cleaning during the preliminary phase of this project, and Clara Medina for their expertise and input on reporting CBPR methods. Field work and community-based research were primarily supported by the Woods Institute at Stanford (Healthy People, Healthy Planet) and a US Environmental Protection Agency Science to Achieve Results (STAR) grant (No. R840478). Additional support was provided by a National Science Foundation (NSF) Graduate Research Fellowship (GRF) to A.G.H., and a Diversifying Academia Recruiting Excellence (DARE) fellowship to L.M.G. from Stanford's Vice Provost for Graduate Education (VPGE) office.

**REFERENCES**

1. Progress on household drinking water, sanitation and hygiene 2000-2022: special focus on gender | JMP. https://washdata.org/reports/jmp-2023-wash-households.

2. Improved sanitation facilities and drinking-water sources. https://www.who.int/data/nutrition/nlis/info/improved-sanitation-facilities-and-drinking-water-sources.

3. Deitz, S. & Meehan, K. Plumbing Poverty: Mapping Hot Spots of Racial and Geographic Inequality in U.S. Household Water Insecurity. *Ann. Am. Assoc. Geogr.* **109**, 1092–1109 (2019).

4. Gasteyer, S. P., Lai, J., Tucker, B., Carrera, J. & Moss, J. BASICS INEQUALITY: Race and Access to Complete Plumbing Facilities in the United States. *Bois Rev. Soc. Sci. Res. Race* **13**, 305–325 (2016).

5. Mueller, J. T. & Gasteyer, S. The widespread and unjust drinking water and clean water crisis in the United States. *Nat. Commun.* **12**, 3544 (2021).

6. He, J., Dougherty, M., Zellmer, R. & Martin, G. Assessing the Status of Onsite Wastewater Treatment Systems in the Alabama Black Belt Soil Area. *Environ. Eng. Sci.* **28**, 693–699 (2011).

7. Carrera, J. S. & Flowers, C. C. Sanitation Inequity and the Cumulative Effects of Racism in Colorblind Public Health Policies. *Am. J. Econ. Sociol.* **77**, 941–966 (2018).

8. Flowers, C. C. Opinion | Mold, Possums and Pools of Sewage: No One Should Have to Live Like This. *The New York Times* (2020).

9. Flowers, C. C. *Waste: One Woman's Fight Against America's Dirty Secret*. (The New Press, La Vergne, 2020).

10. United States Department of Justice. *Interim Resolution Agreement Between the United States and the Alabama Department of Public Health*.

https://www.justice.gov/archives/opa/pr/departments-justice-and-health-and-human-services-announce-interim-resolution-agreement (2023).

11. McKenna, M. L. *et al.* Human Intestinal Parasite Burden and Poor Sanitation in Rural Alabama. *Am. J. Trop. Med. Hyg.* **97**, 1623–1628 (2017).

12. Capone, D. *et al.* Risk Factors for Enteric Pathogen Exposure among Children in Black Belt Region of Alabama, USA. *Emerg. Infect. Dis.* **29**, 2433–2441 (2023).

13. Poole, C. *et al.* Cross-Sectional Study of Soil-Transmitted Helminthiases in Black Belt Region of Alabama, USA. *Emerg. Infect. Dis.* **29**, 2461–2470 (2023).

14. Izenberg, M., Johns-Yost, O., Johnson, P. D. & Brown, J. Nocturnal Convenience: The Problem of Securing Universal Sanitation Access in Alabama's Black Belt. *Environ. Justice* **6**, 200–205 (2013).

15. Maxcy-Brown, J. *et al.* The Past, Present, and Future of Wastewater Management in Alabama's Black Belt. *J. Sustain. Water Built Environ.* **10**, 04024007 (2024).

16. Maxcy-Brown, J. *et al.* Making waves: Right in our backyard- surface discharge of untreated wastewater from homes in the United States. *Water Res.* **190**, 116647 (2021).

17. Bakchan, A. & White, K. D. Identifying Socio-Technical Challenges to Decentralized Wastewater Infrastructure Management in the Rural Alabama Black Belt. 690–700 (2024) doi:10.1061/9780784485279.069.

18. Sanitation | JMP. https://washdata.org/monitoring/sanitation.

19. *Encyclopedia of Measurement and Statistics*. (SAGE Publications, Thousand Oaks, Calif, 2007).

20. US EPA, O. How to Care for Your Septic System. https://www.epa.gov/septic/how-care-your-septic-system (2015).

21. Septic Tank Maintenance | Alabama Department of Public Health (ADPH). https://www.alabamapublichealth.gov/onsite/maintenance.html.

22. Mendoza Grijalva, L., Brown, B., Cauble, A. & Tarpeh, W. A. Diurnal Variability of SARS-CoV-2 RNA Concentrations in Hourly Grab Samples of Wastewater Influent during Low COVID-19 Incidence. *ACS EST Water* acsestwater.2c00061 (2022) doi:10.1021/acsestwater.2c00061.

23. *Wastewater Engineering: Treatment and Resource Recovery*. (McGraw-Hill Education, New York, NY, 2014).

24. *Wastewater Engineering: Treatment and Reuse*. (McGraw-Hill, Boston, Mass., 2003).

25. Davis, M. L. & Cornwell, D. A. *Introduction to Environmental Engineering*. (McGraw-Hill, New York, 2013).

26. U.S. Environmental Protection Agency (EPA). Primer for Municipal Wastewater Treatment Systems. (2004).

27. Maxcy-Brown, J., Capone, D. & Elliott, M. A. Characterizing the nature and extent of access to unsafely managed sanitation in the United States. *Nat. Water* **1**, 915–928 (2023).

28. Sorenson, S. B., Morssink, C. & Campos, P. A. Safe access to safe water in low income countries: water fetching in current times. *Soc. Sci. Med. 1982* **72**, 1522–1526 (2011).

29. Hall, R. P. *et al.* Impact Evaluation of the Mozambique Rural Water Supply Activity. (2014).

30. Paulos, A. *et al.* Temperature and precipitation affect the water fetching time burden in sub-Saharan Africa. Preprint at https://doi.org/10.21203/rs.3.rs-3789072/v1 (2023).

31. Cherukumilli, K., Ray, I. & Pickering, A. J. Evaluating the hidden costs of drinking water treatment technologies. *Nat. Water* **1**, 319–327 (2023).

32. Mills, F. *et al.* Indicators to complement global monitoring of safely managed on-site sanitation to understand health risks. *Npj Clean Water* **7**, 58 (2024).

33. Lebu, S. *et al.* Indicators for evaluating shared sanitation quality: a systematic review and recommendations for sanitation monitoring. *Npj Clean Water* **7**, 102 (2024).

34. Hendricks, M. D. & Rosenberg Goldstein, R. E. Sanitary sewer overflows, household sewage backups, and antibiotic-resistant bacteria: the new frontier of environmental health risks and disasters. *Environ. Res. Health* **3**, 013001 (2025).

35. Oza, H. H. *et al.* Occupational health outcomes among sanitation workers: A systematic review and meta-analysis. *Int. J. Hyg. Environ. Health* **240**, 113907 (2022).

36. Tashakkori, A. & Creswell, J. W. Editorial: The New Era of Mixed Methods. *J. Mix. Methods Res.* **1**, 3–7 (2007).

37. Curry, L. A., Nembhard, I. M. & Bradley, E. H. Qualitative and Mixed Methods Provide Unique Contributions to Outcomes Research. *Circulation* **119**, 1442–1452 (2009).

38. McHugh, M. L. Interrater reliability: the kappa statistic. *Biochem. Medica* **22**, 276–282 (2012).